# Impact of Building-Level Motor Protection on Power System Transient Behaviors

Yuan Liu,[1] Yu Zhang,[1] Qiuhua Huang,[1] Soumya Kundu,[1] Yingying Tang,[1]
Daniel James,[1] Pavel Etingov,[1] Bhaskar Mitra,[1] David P Chassin[2]

1, Pacific Northwest National Laboratory
Richland, WA, USA, 99354

2, SLAC National Accelerator Laboratory
2575 Sand Hill Road, Menlo Park, CA 94025

{yuan.liu, yu.zhang,qiuhua.huang, soumya.kundu,yingying.tang, daniel.james, pavel.etingov}@pnnl.gov, dchassin@slac.stanford.edu

*Abstract*—**Protection strategies for transmission and distribution systems have been extensively investigated to facilitate better coordination of physical protection devices. A diverse range of functional motors with dedicated protection schemes are being used more and more in commercial, residential, and industrial buildings. This paper focuses on simulating several of the most popular protection schemes using the Electro-Magnetic Transient Program (EMTP) model for three-phase and single-phase induction motors in existing commercial buildings connected to typical distribution feeders. To investigate the behaviors of single-phase motors stalling, the actions of motor protection and reconnections, and the impacts of device-level protection on system-level dynamics, we imposed voltage depressions at the head of a feeder fully loaded with functional induction motors. Several distribution feeders are represented in a standard IEEE 39-bus transmission system to simulate fault-induced delayed voltage recovery (FIDVR) and explore mitigation strategies by optimally configuring the building-level motor protection settings.**

## I. INTRODUCTION

Accurate modeling of loads in transmission and distribution systems has increasing importance in power system dynamics simulations. Traditional static load models no longer provide reliable simulation results as the number of dynamic motor loads present in the system increases [1]. In transmission simulation programs such as PSLF and PSS/E, the state-of-the-art WECC composite load model (CMPLDW in PLSF; CMLDBLU1 in PSS/E) [2] contains four motor types, representing the homogeneous characteristics of a group of motors for each of the type. However, this highly aggregated load model cannot fully address the problems arising from the distributed motor loads along the feeder. For example, recent investigations reveal that even motors with similar loadings will have different stalling behaviors because those motors are geographically distant from each other along the feeder. Stalling of upstream motors also is likely to aggravate the stalling of downstream motors along the feeder [3]. Independent protection schemes are used for each of the four types of motors in the CMPLDW model to emulate the aggregated protection responses of a cluster of homogeneous motors. The underlying presumption for using the CMPLDW model and protection schemes is that a fraction of the same categories of motor loads either ride through or disconnect by a specific fault simultaneously. Sensitivity studies conducted by Zhang and Zhu [4] have demonstrated that the composition percentages of the four motor types, along with the protection parameters of each type in CMPLDW, are dominating factors that affect the dynamic responses of the CMPLDW model, particularly during the fault-induced delayed voltage recovery (FIDVR) events. Therefore, the importance of calibrating the composition and protection parameters in CMPLDW has been recognized. Previously, the motor composition and protection parameters were evaluated based on engineering judgement. Recently, some research efforts have focused on developing an accurate motor composition model [5] and investigating realistic motor protection types and parameters [6]. A research paper by Kueck et al. [7] also reported that motor loads in the CMPLDW model have varying sensitivities to different voltage contingencies and types of protections used in the motors.

Accurately parameterizing motor protection models is becoming more and more significant in calibrating the composite load model. To generalize the aggregated protection response of motor loads, calibration of protection parameters must commence with a bottom-up approach. The intent of this paper is to provide implementation guidelines for several critical motor protection models at the distribution end-use level and to investigate the impacts of end-use motor protections on system transient behaviors. The heterogeneous collection of motors present in the system is represented for several typical commercial buildings connected to different nodes of a distribution feeder. Each of the commercial buildings contains a combination of motor loads and ZIP loads. The composition and protection parameters of the heterogeneous motors in each building are based on reviews of previous research publications. Several representative motor protection schemes are described and implemented in detail. The feeder model, motor composition model, and five types of protection models with examined parameters are explicitly developed in the PSCAD simulator. The dynamics of the realistically modeled feeder are observed in the PSCAD simulation by feeding various voltage contingencies at the head of the feeder. Transmission and distribution (T&D) co-simulations are

This work was supported by the U.S. Department of Energy (DOE) Office of Electricity Delivery & Energy Reliability as part of Advanced Grid Research & Development Program. The authors gratefully thank Mr. Ali Ghassemian from DOE for his continuing support, help and guidance performed intentionally to explore the impacts of individual motor protections on large-scale system dynamics.

The paper is organized as follows. Section II provides a discussion of the motor categorization and protection implementation. Section III will present the simulation results using the developed motor loading and protection models. Section IV will conclude the paper.

## II. MODEL DEVELOPMENT

This section first provides basic background information about the four types of motors in the CMPLDW model and categorizes building-level motors into the four types. Then, we discuss implementation of the five common protection schemes. Finally, the feeder formulation is illustrated.

### A. Categorization of Motor Loads

The common types of three-phase or single-phase induction motors found in six typical commercial buildings are investigated through reviews of previously reported work [5], [6]. These six buildings include two retail stores (medium and large), a supermarket, a hospital, a hotel, and a warehouse. The identified motors in these buildings are categorized into the four CMPLDW motor types based on inertia and torque characteristics. The four motor types in WECC composite load model, referred to as Motors A, B, C, and D, are summarized below [2]:

- Motor A: Three-phase induction motors operating under constant torque. Examples include motors used in commercial air-conditioners and refrigerators.
- Motor B: Three-phase induction motors with high inertia operating under speed-dependent torque. Examples include motors used in fans.
- Motor C: Three-phase induction motors with low inertia operating under speed-dependent torque. Examples include motors used in pumps.
- Motor D: Single-phase induction motors. Examples include motors used in air-conditioners and heat pumps.

The ratings of the identified motors were determined from reviews of DOE-conducted survey reports, including an Energy Information Administration commercial building survey [8] and DOE commercial prototype building models [9] used in EnergyPlus simulations [10]. For motor rated data that was not found in DOE reports, a common calculation [11] was applied to find the typical rated motor load for each particular building set. If specified motors fell outside the common calculations, a top-down approach [9] was used to roughly estimate the rated motor load within that building set.

### B. Five Types of Protection

Motor protections are categorized into 5 typical types. The five protection types commonly applied in these identified motors are modeled below:

- Protection 1 (P1): Electronic Relay. This protection trips the motor operating in under-voltage conditions at the terminal. This protection is usually accompanied with reconnection logics.
- Protection 2 (P2): Current Overload Protection. This protection trips the motor if the motor terminal current exceeds a threshold and lasts for a delayed period of time.
- Protection 3 (P3): Thermal Protection. This protection trips the motor when the stator temperature reaches a threshold. This type of protection is widely found in single-phase air conditioner motors.
- Protection 4 (P4): Contactor. This protection trips the motor running in extremely low voltage conditions. This type of protection is usually configured for fast response to severe voltage depression conditions. Contactor protection also has reconnection logics.
- Protection 5 (P5): Building Management System (BMS). Testing conducted by the Bonneville Power Administration has shown that the BMS can ride though severe voltage sags down to 65% of nominal voltage. BMS controllers have reconnection logics.

The motor loads and associated protections for the six buildings are summarized in TABLE I. According to the investigation, some identified motors are equipped with more than one types of protection. From the above descriptions of the protection types, we noted that protections P1, P4, and P5 are triggered by transient low voltage conditions, and each of them has a reconnection logic. These three voltage-dependent protections can be defined identically and implemented with different parameter settings. The current overload and thermal protections are implemented individually. The development of the five protection logics is accomplished in PSCAD using the master library components and user-defined models.

### C. Voltage-Dependent Protection Schemes

The voltage-dependent protection schemes, as discussed in Section II.B, include electronic relay (P1), contactor (P4), and BMS (P5). These three protection logics are defined in the same subroutine of the PSCAD user-defined component with options of enabling or disabling each type. Each protection logic receives the same voltage signal from the motor terminal sensor. If the voltage drops below the trip voltage level of a specific protection, an individual timer will begin to count the length of time the voltage stays below the trip level. If the voltage recovers sooner than the delayed time is reached, the motor does not trip, and the timer would be reset to zero. If the voltage does not recover, a trip signal will be sent. The reconnection for this specific protection type follows similar rules under the pre-conditions that the motor has been disconnected by this protection and voltage recovers above the reconnection threshold. Because some motors have more than one voltage-dependent protection, the outputs of these three protection controllers are logically connected in parallel as inputs of an OR gate to ensure that the final output of the OR gate will be based on a "first-come–first-trip" mechanism.

TABLE I. STATIC LOADS, MOTOR LOADS AND ASSOCIATED PROTECTIONS

| Building | Appliance | Equipment | Motor Type | Protections | Rating (kW) |
|---|---|---|---|---|---|
| Medium Retail | RTU | Fan | MB | P2P4P5 | 15.38 |
| | RTU | Compressor | MA | P2P4P5 | 53.13 |
| | RTU | Frac. Condenser | MD | P3P4P5 | 16.25 |
| | RTU | Frac. Ind. Draft | MD | P3P4P5 | 10.41 |
| | Exhaust | Frac. Fan | MD | P3P4P5 | 0.92 |
| | Static Loads | | | | 41.18 |
| Large Retail | RTU | Fan | MB | P2P4P5 | 46.15 |
| | RTU | Compressor | MA | P2P4P5 | 159.38 |
| | RTU | Frac. Condenser | MD | P3P4P5 | 48.75 |
| | RTU | Frac. Ind. Draft | MD | P3P4P5 | 31.22 |
| | Exhaust | Frac. Fan | MD | P3P4P5 | 1.38 |
| | Static Loads | | | | 122.95 |
| Supermarket | RF | Compressor | MA | P2P4 | 42.5 |
| | RF | Frac. Fan | MD | P3 | 17 |
| | Exhaust | Frac. Fan | MD | P3P4P5 | 1.38 |
| | RTU | Fan | MB | P2P4P5 | 30.77 |
| | RTU | Compressor | MA | P2P4P5 | 106.25 |
| | RTU | Frac. Condenser | MD | P3P4P5 | 32.5 |
| | RTU | Frac. Ind. Draft | MD | P3P4P5 | 20.81 |
| | Static Loads | | | | 107.66 |
| Warehouse | Gas_Heater | Fan | MD | P3P4 | 1.2 |
| | Exhaust | Frac. Fan | MD | P3P4 | 24.62 |
| | Static Loads | | | | 11.07 |
| School | Chiller | Compressor | MA | P1P4P5 | 350 |
| | Chiller | Pump | MC | P2P5 | 98 |
| | Cool_Tower | Fan | MB | P2P4P5 | 42 |
| | Fan_Coil | Fan | MB | P4P5 | 6.15 |
| | Exhaust | Fan | MB | P2P4P5 | 1.29 |
| | Boilers | Ind. Draft | MB | P1P4P5 | 83.25 |
| | Boilers | Pump | MC | P2P5 | 98 |
| | RTU | Fan | MB | P2P4P5 | 123 |
| | RTU | Compressor | MA | P2P4P5 | 425 |
| | RTU | Frac. Condenser | MD | P3P4P5 | 130 |
| | RTU | Frac. Ind. Draft | MD | P3P4P5 | 83.25 |
| | Static Loads | | | | 617.12 |
| Hotel | PTAC | Compressor | MA | P4 | 425 |
| | PTAC | Fan | MD | P3 | 123 |
| | Exhaust | Fan | MD | P3 | 23 |
| | HWP | Pump | MD | P3 | 1.2 |
| | Split | Fan | MB | P2P4 | 123 |
| | Split | Compressor | MA | P2P4 | 425 |
| | Split | Frac. Condenser | MD | P3P4 | 130 |
| | Split | Frac. Ind. Draft | MD | P3P4 | 83.25 |
| | Static Loads | | | | 571.48 |
| Static | MA | MB | MC | MD | Total |
| 1471.45 | 1986.26 | 470.99 | 196.00 | 780.14 | 4904.84 |
| 30.00% | 40.50% | 9.60% | 4.00% | 15.91% | 100.00% |
| Note: Frac. → Fractional, Ind. → Induced | | | | | |

The implementation algorithm of the voltage-dependent protection scheme is outlined in TABLE II.

### D. Current Overload Protection

Current overload protection is implemented using similar tripping logic (lines 13–25 in TABLE II) to the voltage-dependent protection scheme presented in TABLE II. The current measurement ($I_{measured}$) is sent in as an input and compared with trip current threshold ($I_{tr}$). The motor is tripped by the current overload protection after a time delay if $I_{measured} > I_{tr}$. The difference is that there is no reconnection logic for overload protection. Thus, the maximum allowed tripping number (*MaxTripCount*) is hard coded to be 1.

TABLE II. ALGORITHM OF VOLTAGE DEPENDENT PROTECTION

**Algorithm 1** Voltage Dependent Protection
1: **Subroutine** MyProtection
2: *ProtActivated* ← True or False (user specified)
3: *ProtWorkTime* ← time protection begins to work (user specified)
4: *Vtr* ← trip voltage (user specified)
5: *Ttr* ← trip delay (user specified)
6: *Vrec* ← reconnection voltage (user specified)
7: *Trec* ← reconnection delay (user specified)
8: *MaxTripCount* ← maximum allowed tripping count (user specified)
9: *Initialize Variables:*
10: *Time* ← 0.0, *TripTimer* ← 0.0, *RecTimer* ← 0.0, *TripCounter* ← 0
11: *Loop:*
12: **If** (*ProtActivated* = True) AND (*Time* >= *ProtWorkTime*) **Then**
13:   **If** *Vmeasured* < *Vtr* **Then**  //Tripping logic initiated
14:     *TripTimer* ← *TripTimer* + Δ*t*
15:     **If** *TripTimer* > *Ttr* **Then**
16:       **If** *ProtTrip* = False **Then**
17:         *TripCounter* ← *TripCounter* + 1
18:       **End if**
19:       *ProtTrip* ← True
20:     **Else**
21:       *ProtTrip* ← False
22:     **End if**
23:   **Else**
24:     *TripTimer* ← 0.0
25:   **End if**
26:   **If** *ProtTrip* = True **Then**  //Reconnection logic initiated
27:     **If** *Vmeasured* > *Vrec* **Then**
28:       *RecTimer* ← *RecTimer* + Δ*t*
29:       **If** *RecTimer* > *Trec* **Then**
30:         *ProtTrip* ← False
31:         *TripTimer* ← 0.0
32:       **End if**
33:     **Else**
34:       *RecTimer* ← 0.0
35:     **End if**
36:   **End if**
37:   **If** *TripCounter* >= *MaxTripCount* **Then**  //Check trip counter
38:     *ProtTrip* ← True
39:   **End if**
40: **End if**
41: *Time* ← *Time* + Δ*t*, **goto** *Loop*
42: **Output:** *ProtTrip*

### E. Thermal Protection

The standard thermal protection model used in the PSLF performance-based model of an air-conditioner (ld1pac) [2], [12] is implemented for all the single-phase motors identified as MD in TABLE I.

The thermal protection logic, which is connected to the third input port of the OR gate, is shown in Fig. 1. When the motor is stalled, the current drawn by the stalled motor is represented by a constant impedance load ($R_{stall} + jX_{stall}$). The temperature of the motor is computed by integrating $R_{stall}I^2_{measured}$ through the thermal time constant $T_{therm}$ in the first-order transfer function in Fig. 1. The integrated result, which represents the motor temperature, is compared with a threshold temperature $T_{th}$. The motor is tripped when the motor temperature exceeds the threshold.

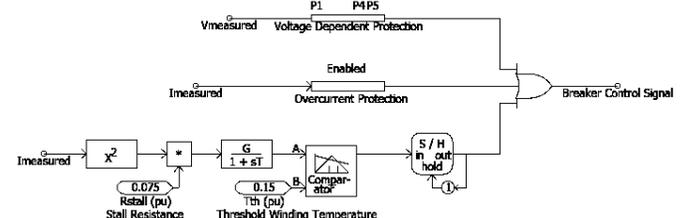

Fig. 1. Schematic of protection logics

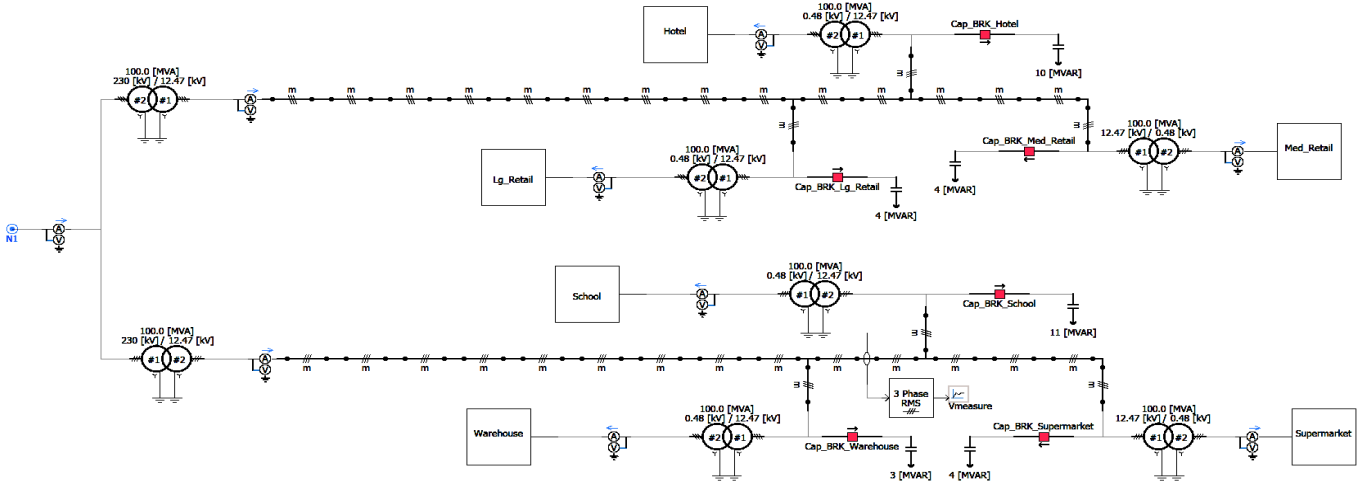

Fig. 2 Distribution Feeder Schematic

## F. Capacitor Bank Over-Voltage Tripping

A capacitor bank is deployed at the high-voltage side of each building transformer to provide VAr support. The over-voltage tripping mechanism is implemented for these capacitor banks during a cascaded motor tripping event caused by voltage depression. The status of the capacitor bank does not change if the terminal operating voltage $V_{op}$ stays within a range. The capacitor bank is tripped when $V_{op}$ rises above the upper bound, and is reconnected when $V_{op}$ drops below the lower bound. The mechanism is expressed by (1)–(4).

$$Initialize: Status \leftarrow on \quad (1)$$
$$if\ V_{op} \geq V_{max}\ then, Status \leftarrow on \quad (2)$$
$$else\ if\ V_{min} < V_{op} < V_{max}\ then, Status \leftarrow Not\ Change \quad (3)$$
$$else, Status \leftarrow off \quad (4)$$

## III. STUDY CASES

In this study, the six commercial buildings are supplied by the distribution taxonomy feeder GC-12.47-1 [9], as shown in Fig. 2. The feeder consists of 30% static (ZIP) loads and 70% motor loads. The percentage of each motor type is shown in TABLE I. This feeder is connected to Bus 18 of the IEEE 39 bus system [13] to replace the original ZIP load, as illustrated by Fig. 3.

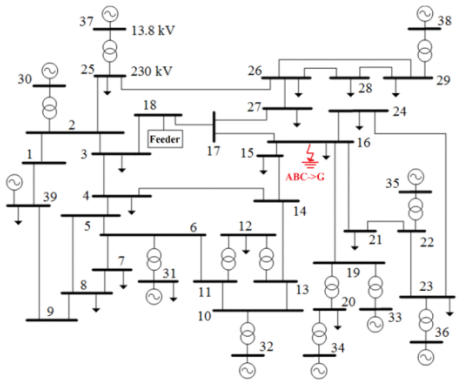

Fig. 3 IEEE 39 bus transmission system [13]

The building load ratings in TABLE I are identically scaled to match the original load at Bus 18. A six-cycle three-phase-to-ground fault is applied at Bus 16 of the transmission system, depression the voltage at Bus 18 to 0.35 pu. The simulation considers two scenarios: Scenario A when all of the described protections for each motor in TABLE I are activated and Scenario B when only the thermal protection (P3) is activated and the other protections are deactivated. The parameters of the five protections used in the simulation are generated from the specific numbers or randomly picked up from the ranges described below [6], [7]:

- P1: $V_{tr} = 0.8 \sim 0.9\ pu$, $T_{tr} = 20\ cycle \sim 2.0\ s$, $V_{rec} = 0.95\ pu$, $T_{rec} = 0.01\ s$
- P2: $I_{tr} = 3.0\ pu$, $T_{tr} = 0.04\ s$
- P3: $T_{th} = 0.15\ pu$, $T_{therm} = 10\ s$, $R_{stall} = 0.054 \sim 0.086\ pu$
- P4: $V_{tr} = 0.4 \sim 0.6\ pu$, $T_{tr} = 1 \sim 5\ cycle$, $V_{rec} = 0.65 \sim 0.7\ pu$, $T_{rec} = 2 \sim 8.5\ cycle$
- P5: $V_{tr} = 0.5 \sim 0.6\ pu$, $T_{tr} = 13 \sim 15\ cycle$, $V_{rec} = 0.95\ pu$, $T_{rec} = 2.0\ s$

Fig. 4 shows the voltage at Bus 18, the head of the feeder. It can be seen that with all protections activated, the post-event voltage is a little higher than the scenario in which only thermal protection is enabled because some motors are tripped offline by fast-reacting current overload or contactor protections (P2 and P4).

The voltage between the warehouse and school buildings is measured by a 3-phase RMS voltage meter, as shown in Fig. 2. The measured voltages in Scenarios A and B are compared in Fig. 5 (A). The voltage bump observed at t = 2.2 s in Fig. 5 (A)(B) is caused by the tripping and reconnection of a large chiller compressor in the school building. The maximum allowed tripping number (*MaxTripCount*) for electronic relay (P1) is set to be 2, meaning no reconnection action after the second tripping. The action sequence of chiller compressor protections is illustrated in Fig. 5 (B).

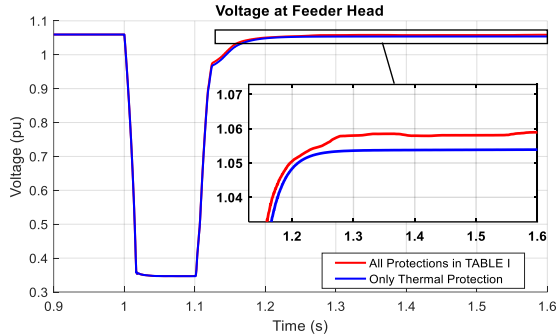

Fig 4. Voltage at Bus 18 (Feeder head voltage)

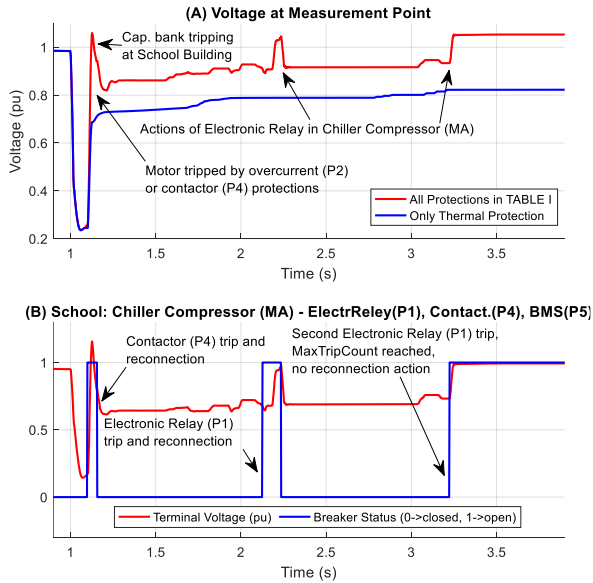

Fig. 5. Voltage at measurement point and plots of a large chiller compressor in a school building

Fig. 6 shows the performance of coordinated overload (P2) and contactor (P4) protections. The roof-top unit (RTU) fan motor is first tripped by contactor. As voltage recovers, the contactor recloses and the motor reaccelerates. The current drawn during reacceleration exceeds the threshold current of P2 at 1.222 s. The motor is irreversibly tripped by the overload protection at 1.262 s.

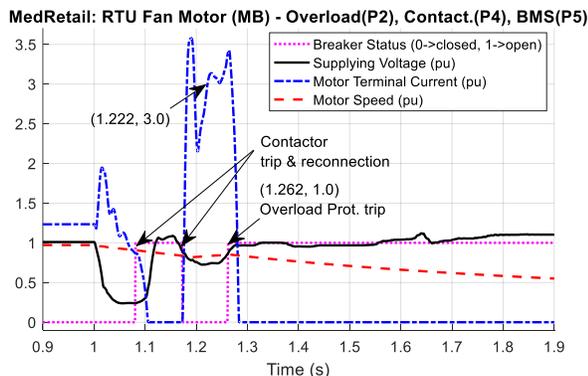

Fig. 6. Performances of overload protection (P2) and contactor (P4) of RTU fan motor (MB) in a medium-size retail building

Fig. 7 shows the coordinated behaviors of thermal (P3) and contactor (P4) protections.

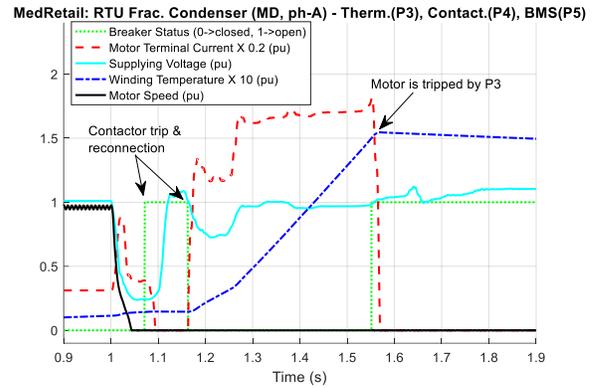

Fig. 7. Performances of thermal protection (P3) and contactor (P4) of chiller pump motor (MC) in a school building

IV. CONCLUSIONS

In this paper, the impacts of end-use motor loads with protections on power system transient behaviors are studied. The motor loads and their protections in six types of commercial buildings are categorized and modeled in EMTP T&D co-simulations. IEEE 39 bus system and a typical distribution taxonomy feeder are used in the PSCAD simulation. The realistic system dynamics can be properly simulated by modeling the motor loads and protections in detail.